\documentclass[conference]{IEEEtran}
\IEEEoverridecommandlockouts
\usepackage{cite}
\usepackage{amsmath,amssymb,amsfonts}
\usepackage{algorithmic}
\usepackage{graphicx}
\usepackage{textcomp}
\usepackage{xcolor}
\usepackage{listings}
\usepackage{times}
\usepackage{fancyhdr,graphicx,amsmath,amssymb}
\usepackage[ruled,vlined]{algorithm2e}
\usepackage{tikz-qtree}
\usepackage{graphicx}
\usepackage{enumitem}
\usepackage{float}

\graphicspath{ {./images/} }
\include{pythonlisting}
\def\BibTeX{{\rm B\kern-.05em{\sc i\kern-.025em b}\kern-.08em
    T\kern-.1667em\lower.7ex\hbox{E}\kern-.125emX}}

\begin{document}

\title{Improving a High Productivity Data Analytics Chapel Framework}

\author{\IEEEauthorblockN{Prashanth Pai}
\IEEEauthorblockA{\textit{Rice University} \\
pbp2@rice.edu}
\and
\IEEEauthorblockN{Andrej Jakovljević}
\IEEEauthorblockA{\textit{University of Belgrade} \\
ja180039d@student.etf.bg.ac.rs}
\and
\IEEEauthorblockN{Zoran Budimlić}
\IEEEauthorblockA{\textit{Rice University} \\
zoran@rice.edu}
\and
\IEEEauthorblockN{Costin Iancu}
\IEEEauthorblockA{\textit{Lawrence Berkeley National Laboratory} \\
cciancu@lbl.gov}
}

\maketitle

\begin{abstract}

Most state of the art exploratory data analysis frameworks fall into one of the two extremes: they either focus on the high-performance computational, or on the interactive and open-ended aspects of the analysis.  Arkouda is a framework that attempts to integrate the interactive approach with the high performance computation by using a novel client-server architecture, with a Python interpreter on the client side for the interactions with the scientist and a Chapel server for performing the demanding high-performance computations. The Arkouda Python interpreter overloads the Python operators and transforms them into messages to the Chapel server that performs the actual computation.

In this paper, we are proposing several client-side optimization techniques for the Arkouda framework that maintain the interactive nature of the Arkouda framework, but at the same time significantly improve the performance of the programs that perform operations running on the high-performance Chapel server. We do this by intercepting the Python operations in the interpreter, and delaying their execution until the user requires the data, or we fill out the instruction buffer. We implement caching and reuse of the Arkouda arrays on the Chapel server side (thus saving on the allocation, initialization and deallocation of the Chapel arrays), tracking and caching the results of function calls on the Arkouda arrays (thus avoiding repeated computation) and reusing the results of array operations by performing common subexpression elimination. 

We evaluate our approach on several Arkouda benchmarks and a large collection of real-world and synthetic data inputs and show significant performance improvements between 20\% and 120\% across the board, while fully maintaining the interactive nature of the Arkouda framework. 

\end{abstract}

\begin{IEEEkeywords}
distributed processing; parallel programming; client-side optimization; caching;
Arkouda; triangle counting; betweenness centrality
\end{IEEEkeywords}

\section{Introduction}

In data science, exploratory data analysis (EDA) is a very common approach, with the scientist interacting with the data and performing several complex operations on it, to attempt to glean some insight into the meaning of the data. The operations that such EDA applications perform are often very complex and computationally intensive, so it would be often desirable to perform those operations on a high-performance computing platform. Unfortunately, the interactive nature of EDA does not lend itself easily to an execution on a high-performance platform, since EDA is most of the time done on the scientist's laptop or other low-performance platform.

Arkouda~\cite{Arkouda,ArkoudaGitHub} bridges this gap by implementing a client-server framework, where the client is implemented as a Python interpreter that can run on a low-performance machine (such as the scientist's laptop), while the server is implemented in Chapel and can run on high-performance distributed platforms. The client uses a simple, NumPy-like API to express operations on the data, turns those operations into messages to the server, and the server processes these messages and executes the corresponding data operations on the actual data on the server. All the data resides on the server, until the scientist requests some of it (by, for example, plotting the data, querying it, reducing it to a value, or extracting some part of it), at which point the server transfers the requested part of the data to the client. The expectation is that in most of the EDA applications, the amount of data requested by the client will be significantly smaller than the total amount of data residing on the server.

However, Arkouda implements this framework by immediately translating Arkouda Python commands on the client into messages to the server to perform the actual server actions, even if the commands do not result in anything that is immediately visible to the user. This could potentially lead to many unnecessary server operations, and significantly reduce the available throughput for the EDA application.

In this paper, we propose several optimizations to the Arkouda framework that take advantage of the fact that the results of many server-side operations are not immediately requested by the user, and their execution could be delayed up to the point when the user actually requests to see the results. For example, in the statement $A = B * C$, where $A$, $B$ and $C$ are Arkouda arrays, the multiplication does not need to be performed until the user requests to see the data that $A$ contains (by, for example, plotting it, or finding a maximum, or indexing into it). If the arrays $A$, $B$ or $C$ are used in other computations in the meantime, this provides an opportunity to the client to delay the computation, reorganize the server-side operations, and reduce the necessary server-side computations needed to provide the user with the requested data. 

This client-server architecture presents several unique challenges to compiler optimizations. First, the optimizations have to be fast, since they will be executed at runtime. Second, the interactive nature of the Arkouda framework has to be maintained. Finally, the client and the server need to be in agreement on the changes to their respective internal representations. 

The remaining of the paper is organized as follows: In Section~\ref{sec:Related}, we discuss the related work, while in Section~\ref{sec:Arkouda}, we describe the architecture of the Arkouda framework. In Section~\ref{sec:Model} we propose the changes to the client and the server to enable our optimizations. Section~\ref{sec:Examples} shows a walk-through of several examples of how these optimizations work. Section~\ref{sec:Experiments} presents several benchmarks, and the experimental evaluation of our approach compared to the base Arkouda. Section~\ref{sec:Future} describes several opportunities for future optimizations, while Section~\ref{sec:Conclusions} concludes the paper.

\section{Related Work}
\label{sec:Related}
Our work builds upon the existing Arkouda~\cite{Arkouda,ArkoudaGitHub} framework, which offers an interactive way to perform parallel data processing on distributed datasets using a Chapel server and a simple Python client. Our project optimizes this client by introducing common subexpression elimination, caching, and lazy evaluation. The client interface exposes a NumPy-like API to the user. NumPy~\cite{NumPy} is a popular data science package that implements multidimensional vectors and supports distributed libraries, providing an effective building block for Arkouda and our work. Rather than explicitly initializing NumPy arrays, Arkouda and our model creates parallel distributed one-dimensional arrays with functions named in a very similar manner to that of NumPy. As explained in the Arkouda documentation, the overall goal of this framework and our proposed model is to scale existing data science frameworks such as Pandas~\cite{Pandas}, a widely used data analysis tool. Arkouda offers a way to operate on large scale Pandas-like dataframes serving as a major breakthrough for data scientists~\cite{ArkoudaGitHub}.

One model that our project resembles is Phoenix~\cite{Phoenix}, a parallel programming model which sends messages between nodes in the effort of fusing compute resources. We implement a similar idea, but using a client-server model. We have also investigated the use of lambdas to create a dynamic experience for users rather than having them rely solely on a predefined API. ~\cite{Pearl} similarly allows for custom computational patterns, by relying on using two functional languages, RISE and ELEVATE, to produce optimized HPC code. Our work also builds off of functional patterns, relying on overloading Python functions and Chapel APIs.

Others have developed projects for improving Chapel-based systems, such as extending the Arkouda server with a Chapel-based version of the triangle counting algorithm~\cite{ArkoudaGraphStreams}. Kayraklioglu et al. ~\cite{ChapelManycore} compare Chapel and OpenMP implementations of The Parallel Research Kernel and optimize the Chapel compiler. Our project assumes an optimized Chapel server, largely influenced by projects similar to the two aforementioned ones, and allows for general algorithm implementations in Python rather than specialized server-side ones.

JAX~\cite{JAX} is a popular just-in-time compiler that improves Python code through the use of optimized kernels and GPUs. Numba~\cite{Numba} is another Python optimization library. It converts Python code on the fly to optimized machine code using LLVM. Our approach uses similar optimization tactics in a unique client-server environment to effectively translate Python code to high-performance Chapel code.

\section{Current Arkouda Framework}
\label{sec:Arkouda}

\subsection{Client-Server Model}

The Arkouda architecture is uniquely split into a Python client and a Chapel-backed server~\cite{Arkouda,ArkoudaGitHub}. The Python client exposes a simple, Numpy-like API to the user. To represent arrays, the Python client relies on a $pdarray$ (parallel distributed array) class as a thin Python proxy for the server-side arrays. By using operator overloading for common array functions such as addition, multiplication, and other binary operations, this approach allows the user to use intuitive Python expressions for array operations without worrying about interfacing with the Arkouda server code. The client transforms the Python $pdarray$ operation into a message to the server via ZeroMQ~\cite{ZeroMQ}, and waits for a reply. Thus, each Arkouda operation is a blocking call. The $pdarray$ class acts as a proxy to the actual arrays that are created, stored and manipulated on the server. $Pdarray$s only contain the key metadata from the server-side arrays, such as an ID, size, and data type. To delete server-side arrays, the client relies on overloading the $pdarray$ destructor, reference counting, and the Python garbage collector. When there is no longer a reference to a $pdarray$, the garbage collector deletes the unused object, which invokes the overloaded class destructor. This, in turn, sends a message to the server to delete the corresponding distributed array. 

On the server side, the Chapel server receives messages, parses the message commands, maps them to Chapel functions, and performs computations in parallel and distributed fashion on the server. These computations operate on distributed arrays. When the command sent to the server results in a scalar, the server responds to the client with the direct result of the corresponding computation. For example, a call to the $sum$ command on an array would result in the server returning a single value to the client. When the server receives a command which results in a new array (for example, adding two arrays), the server creates and stores a new distributed array and responds to the client with the metadata of the resulting array, which in turns causes the client to create a $pdarray$ to store the metadata for the newly created server-side array. In subsequent operations on that array, the client can simply send the ID of the proxy object. While this is a very functional approach that leads to a very simple client-side interpreter implementation, it could lead to many unnecessary array creations, deletions and computations on the server side.

\begin{figure}
\centerline{\includegraphics[width=1.0\linewidth]{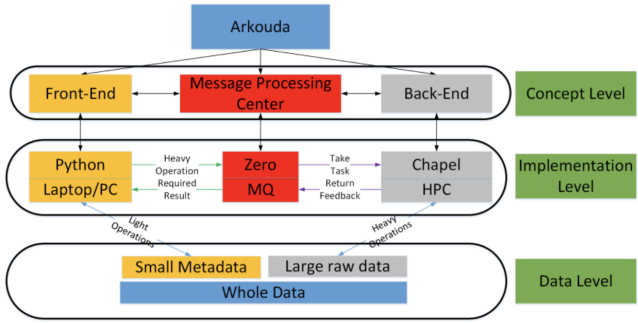}}
\caption{Split between Python client and Chapel server~\cite{ArkoudaGraphStreams}}
\label{fig:PythonChapelArchitecture}
\vspace{-.1in}
\end{figure}

Figure~\ref{fig:PythonChapelArchitecture} shows the general Arkouda client-server architecture. This design allows for a clear separation of function between the client and server. The client acts as a lightweight tool with a proxy class that overloads key functions, and can be executed on low performance platforms, such as a scientist's laptop. The server, on the other hand, handles heavy computations and data storage. This paper focuses on optimizing the client side of the Arkouda framework to reduce the workload on the server.

\subsection{Optimization opportunities}

We propose integration of several client-side optimizations such as lazy evaluation, common subexpression elimination, and array caching into this unique Arkouda framework. When combined, these optimizations reduce command and message traffic, the number of array allocations, and the computational overhead of operating on distributed arrays. To better understand what can be optimized, let us walk through a simple Arkouda example. Assume we have the piece of Python code from Figure~\ref{fig:ArkoudaSampleCode}. Figure~\ref{fig:CurrentObjectAllocation} displays the object allocation scheme that the current Arkouda framework would take for this piece of code. In the figure, client-side Python proxy objects are green and server-side arrays are red.

\begin{figure}
\begin{center}
\begin{tabular}{c}
\begin{lstlisting}
import arkouda as ak
A = ak.randint(0, 10, 10)
B = (A * A) + (A * A)
C = ak.randint(0, 10, 10)
print(B)
\end{lstlisting}
\end{tabular}
\end{center}
\caption{Sample Arkouda code with room for optimization}
\label{fig:ArkoudaSampleCode}
\vspace{-.1in}
\end{figure}

When the second line of code from Figure~\ref{fig:ArkoudaSampleCode} executes, the client would send the server a message to create a server-side one-dimensional distributed array $S1$ of length 10 randomly filled with integers ranging from 0 to 10. The server would use Chapel to create this array and respond to the client with a message containing the metadata for the newly created array. Finally, the client would store these details inside a proxy $pdarray$ object $C1$, and the Python interpreter would assign this object to $A$. 

When the third line of code from Figure~\ref{fig:ArkoudaSampleCode} executes, the Python interpreter would interpret the expression $(A*A)+(A*A)$, which will in turn call the overloaded $pdarray$ operators $*$ and $+$, which will result in a total of 5 messages being sent to the server:
\begin{enumerate}[topsep=0pt,itemsep=-1ex,partopsep=1ex,parsep=1ex,leftmargin=*]
\item First, the client would send a message to the server to multiply $S1$ by itself. This would result in a creation of a new temporary distributed array $S2$ on the server side, and of a proxy $pdarray$ object $C2$ on the client side. 
\item The same process would happen again to complete the rightmost part of the expression which also multiplies A by itself. This would result in a server-side array $S3$ and client-side $pdarray$ object $C3$.
\item Finally a message is sent to the server to add server-side arrays $S2$ and $S3$. This results in another temporary server-side distributed array $S4$, and a  client-side $pdarray$ object $C4$, which is assigned to $B$. 
\item Eventually, two more messages will be sent to the server to delete $S2$ and $S3$ since they are no longer needed to compute $B$. When this is done will depend on the implementation of the Python garbage collector, but most Python implementations will delete temporaries immediately after the statement $B = (A * A) + (A * A)$ is executed.
\end{enumerate}

The fourth line on Figure~\ref{fig:ArkoudaSampleCode} creates another server-side array $S5$ and client-side $pdarray$ proxy object $C5$ assigned to $C$. Finally, the proxy $pdarray$ object $C4$ will be used to create a message that the client sends to the server to retrieve the contents of the distributed array  $S4$. The server would respond to the message with the contents of that array which will be printed to the user. 

\begin{figure}
\centerline{\includegraphics[width=0.7\linewidth]{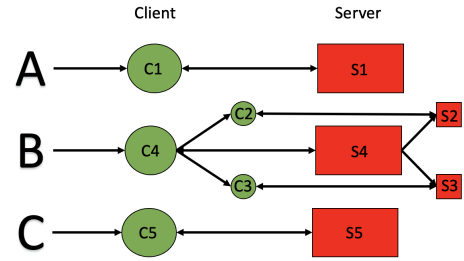}}
\caption{Object allocation of Figure~\ref{fig:ArkoudaSampleCode} code with the current Arkouda framework}
\label{fig:CurrentObjectAllocation}
\vspace{-.1in}
\end{figure}

\begin{figure}
\centerline{\includegraphics[width=0.63\linewidth]{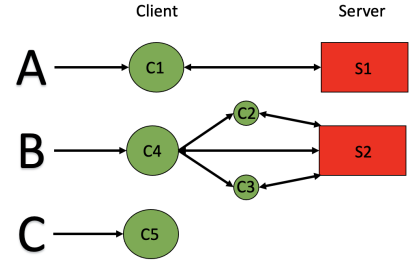}}
\caption{Object allocation of Figure~\ref{fig:ArkoudaSampleCode} code with the optimized Arkouda framework}
\label{fig:OptimizedObjectAllocation}
\vspace{-.1in}
\end{figure}

The code on Figure~\ref{fig:ArkoudaSampleCode} presents several optimization opportunities. First, the computation of $A*A$ does not need to be done twice. Rather than computing both $S2$ and $S3$, we could only compute $S2$ and reuse it. Additionally, this same temporary can be reused to store the result of $S2 + S2$. Finally, since the user is only printing the contents of $B$, which is independent of $C$, we can delay the server call on line 4. Figure~\ref{fig:OptimizedObjectAllocation} shows object allocations in the optimized Arkouda framework.

With these optimizations, the server would only need to create two arrays ($S1$ and $S2$) and respond to four messages (create $S1$, multiply $S1$ with $S1$ and store into new temporary $S2$, add $S2$ and $S2$ and store into $S2$, and retrieve the contents of $S2$). This would be a major improvement over the five arrays and eight messages that the current framework produces. While this is a simple example designed to illustrate the opportunities, in the next section we describe the changes to the client architecture to enable the implementation of these optimizations.

\section{Proposed Model}
\label{sec:Model}

To enable client-side optimizations, we use an abstract syntax tree (AST) based command buffer in the client, memoization to keep track of available expressions and results of reductions, and an extension to the server side API that allows overwriting of existing distributed arrays. Following subsections describe these components and how our optimizations target Arkouda's unique client-server architecture.

\subsection{AST-Based Command Buffer}
To enable on-the-fly analysis and optimization, we introduce a command buffer of abstract syntax trees (AST's) for Arkouda expressions. For example, an expression $$B = (A+A) * (A+A)$$ will be represented by the following binary tree:

\begin{center}
\begin{tabular}{c}
\begin{tikzpicture}
\Tree
[.*     
    [.+ 
       \edge[]; {A}
       \edge[]; {A}
    ]
    [.+ 
       \edge[]; {A}
       \edge[]; {A}
    ]
]
\end{tikzpicture}
\end{tabular}
\vspace{-.1in}
\end{center}

AST's for expressions are represented implicitly, through a use of a command buffer that stores a sequence of Arkouda commands. This  imposes an order amongst commands, and allows for a simple and fast analysis. Additionally, we impose a size constraint on this buffer to avoid unbounded delays in computation. In our model, each pdarray holds a reference to its corresponding buffer command which needs to be executed for the array to have a server-side value. When the value of a $pdarray$ is needed, we search through the buffer for any dependencies and anti dependencies and then execute the corresponding pdarray's buffer command. This naturally leads to the on-demand lazy evaluation of Arkouda expressions.

\subsection{Array Cache}

In our scheme, there is a no longer a one-to-one correspondence between $pdarray$s and server side arrays. Instead, we maintain a cache of server-side arrays that can be reused once their corresponding client-side $pdarray$ is killed by a Python operation. For example, consider the following piece of code:

\begin{center}
\begin{tabular}{c}
\begin{lstlisting}
import arkouda as ak
for x in range (1000):
	A = ak.randint(0, 10, 10)
	B = ak.randint(0, 10, 10)
	C = A + B
	print(C)
\end{lstlisting}
\end{tabular}
\end{center}

For every iteration of the loop, the server creates and later destroys temporaries for $A$, $B$, and $A + B$. However, we could instead use only 3 temporaries rather than 3000. To do so, our model uses a finite cache of server arrays on the client-side. Once the Python interpreter invokes the destructor of a $pdarray$ instance, rather than sending a delete message to the server to delete the server’s instance of that array, we instead cache the server’s id of that array instance. This cache is partitioned by array sizes and data types. When an Arouda command needs to create a server-side array, rather than immediately sending a message to the server to do so, we first check to see if a cached temporary array is available, and if so, use that. We maintain a map of client IDs to server IDs on the client side. This design also allows for a one server with multiple clients framework, with each client with their own virtual space of IDs. We can consequently save on the overhead of constant server array initializations.

\subsection{Command Result Caching}

In addition to array caching, our model caches the results of reductions. Specifically, when the client initiates a reduction function, such as sum, max, or min, on an array, it saves the function result on the client side. By doing so, the server only needs to compute each reduction function up to one time for each array. Additionally, our model caches the results of binary operations and stores those results in a table. Prior to sending a computational message to the server, the client first checks the function cache table to see if the command has already been performed, and uses the cached result if so. When an array is modified, then all the cached expressions transitively using that array are evicted from the cache. Command result caching directly allows for caching of the results of reducing common subexpressions, opening opportunities for common subexpression elimination.

\subsection{Server Side Function Additions}
To allow the reuse of temporary server-side arrays, we extended the server-side APIs with an additional argument to hold the array into which the result should be stored. For example, the Arkouda binary operator command accepts an operator and two array ids as its arguments. The Arkouda server always creates a new array to store the  results of executing the operation. With our improvements, we extend this API with an additional array id to specify where to store the result. This enables the client to send messages to the server which reuse the arrays on the server-side. These "store" functions can be created for any function where using a cached array is possible. 

These functions integrate well with the design of the command buffer. Take for example the following piece of code where $A$, $B$, $C$, and $D$ are Python variables corresponding to $C1$, $C2$, $C3$ and $C4$ Arkouda pdarrays, respectively:

\begin{center}
\begin{tabular}{c}
\begin{lstlisting}
C = B + A
print(C)
C = ak.randint(0, 10, 10)
A = D + A
\end{lstlisting}
\end{tabular}
\end{center}

\noindent Once A is reassigned to $D + A$, the pdarray $C1$ which $A$ used to point to will be up for deletion. However, our model maps client IDs to buffer commands in which that ID was last referenced. In the case above, we can determine that $C1$ can be used to store the result of $D + A$ since it was last referenced in this line. We can then use a "store" function and specify that the server-side array to which $C1$ is mapped to can be used to store the result of $D + A$.

\subsection{Deleted Array Caching}
One potential issue with the model proposed so far is that we can prematurely delete $pdarray$s. For example, assume that we alter the piece of code from the previous example to the following:
\begin{center}
\begin{tabular}{c}
\begin{lstlisting}
C = B + A
A = D + A
print(A)
print(C)
\end{lstlisting}
\end{tabular}
\end{center}

\noindent We cannot simply overwrite the contents of $A$ in line 1 when we reassign $A$ in line 2.

Our model solves this problem by introducing shallow copies of $pdarray$s for any distributed arrays which are still in use at the time of reassignment. These shallow copies hold metadata and instructions, similar to $pdarray$s, but do not use function overloading since the copies will never be used again by the user. In this example, we know the client array ID $C1$ of the array represented by $A$ before $A$ gets reassigned. Since we can traverse the command buffer and validate that $C1$ is still in use in the instruction set for $C$, we can opt to create a shallow copy of $A$ in $A$'s destructor and map $C1$ to that copy. By doing so, when we need to compute the value of $C$, we still have access to the metadata and instruction set of the previous version of $A$. 

\section{Putting it All Together}
\label{sec:Examples}

We will now walk through a few concrete examples of how this architecture works and improves the client-server interaction, in contrast to other just-in-time compilation frameworks. 

\subsection{Example 1}

The first example makes use of every piece of our model, including array caching. Assume we have the following block of Python code where $ak$ refers to the Arkouda library.
\begin{center}
\begin{tabular}{c}
\begin{lstlisting}
A = ak.randint(0, 10, 10)
B = ak.randint(0, 10, 10)
C = ak.randint(0, 10, 10)
C = B + A
A = C + A
print(A)
\end{lstlisting}
\end{tabular}
\end{center}

In Line 1, the client would create a $pdarray$ $C1$ and assign that to $A$. $C1$ would hold onto a reference to a new item in the buffer, which simply encapsulates the instructions of how to create $A$ on the server side. The same process would occur again for Lines 2 and 3, but the client-side IDs for the $pdarray$s assigned to $B$ and $C$ would be $C2$ and $C3$, respectively. 

When Line 4 executes, a fourth $pdarray$ $C4$ and corresponding buffer item would be created and assigned to $C$. The buffer item would hold the information that the newly created $pdarray$ needs: the result of performing a binary operation between $C2$ and $C1$. This buffer item only needs to remember the command name and client side ids of any command inputs.

At this point, $C$ no longer refers to $C3$. Since the reference count to $C3$ would be decremented to 0, the Python interpreter would execute $C3$'s destructor. In most Python implementations, the interpreter invokes the destructor immediately. Using the overloaded pdarray destructor, our model would first traverse through the buffer, processing the most recent entries first, and make sure that no command uses $C3$ as an input. After verifying that $C3$ is not in use anywhere and that its server side value was never computed, our model would remove $C3$'s corresponding buffer item from the buffer and delete $C3$. 

When Line 5 executes, our model creates a fifth $pdarray$ $C5$, adds a corresponding buffer item to the command buffer containing instructions to perform a binary operation on $C4$ and $C1$, and assigns $C5$ to $A$. Similar to before, the Python interpreter would invoke $C1$'s destructor since $A$ is assumed to be reassigned. However, in this case, $C1$ still needs to be used to compute $C4$'s value in Line 4 and $C5$'s value in Line 5. Thus, as detailed in Section~\ref{sec:Model}, we would create a shallow copy of $C1$ for later use and allow $C1$ to be deleted. We would also keep track of where $C1$ is used for the last time, which in this case is in the command to compute $C5$.

When Line 6, executes, our client realizes that it needs to return the value of $C5$ which is only possible through the computation of $C5$'s server side value. Our client would then look at $C5$'s corresponding buffer item which holds the computation instructions for this array, recursively resolve any dependencies and anti-dependencies, and return a value. Since $C5$ is dependent on $C4$ and the shallow copy of $C1$, we would traverse through the buffer and look for any commands that write to $C4$ or $C1$ or read from $C1$. This traversal would first result in the computation of the shallow copy of $C1$ which would involve the client sending a message to the server and the creation of the server side array $S1$. The next relevant buffer item would be the one that contains instructions for computing $C4$. We can see that $C4$ is dependent on $C2$ and the shallow copy of $C1$. The client recursively computes the value of $C2$, using a similar method to the dependency resolving algorithm for $C5$. This would result in the creation of $S2$, the server side representation of $C2$. Since all of $C4$'s dependencies are now resolved, the client then proceeds to compute $S3$, the server side value of $C4$ and the result of adding $S1$ and $S2$. We have now resolved all of $C5$'s dependencies and can overwrite $S1$ to store the server side value of $C5$ and the result of adding $S1$ and $S3$. Our model overwrites $S1$ since we kept track of the information that $C1$ is used for the last time to compute $C5$. Finally, we would ask the server for the contents of $S1$ and print that out to the user.

\begin{figure}
\centerline{\includegraphics[width=0.5\linewidth, height=1.5in]{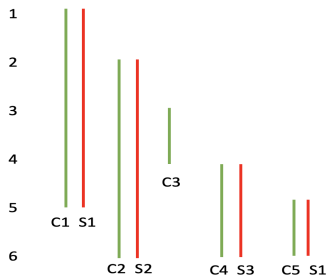}}
\caption{Liveness analysis conducted by our model}
\label{fig:LivenessAnalysis}
\vspace{-.1in}
\end{figure}

Figure~\ref{fig:LivenessAnalysis} illustrates the live ranges of client-side and server-side temporaries from this example. The current Arkouda framework would have created 5 $pdarray$s and 5 server side arrays for this example. While our model also creates 5 $pdarray$s, it only creates 3 server side arrays. 

\subsection{Example 2: Delayed Computations}
This example illustrates the effect of lazy evaluation:
\begin{center}
\begin{tabular}{c}
\begin{lstlisting}
C = B + A
D = E + F
print(D)
\end{lstlisting}
\end{tabular}
\end{center}

Assume $A$, $B$, $E$, and $F$ have been respectively assigned to the client-side $pdarray$ objects $C1$, $C2$, $C3$, and $C4$. Their corresponding arrays have yet to be initialized on the server. The first line of code would result in $C5$, a new $pdarray$ being assigned to $C$ and a corresponding buffer item being placed in the buffer. Similarly, the second line of code would result in $C6$ being assigned to $D$ and a corresponding buffer item being placed in the buffer. When the third line of code executes, since the value of $D$ is needed by the user, our model will find the buffer item that $C6$ points to and execute the instructions stored there. In this case, that would involve computing $C3$ and $C4$'s server side values using their respective buffer items. Once these dependencies have been resolved, our model simply retrieves the contents of the server-side array that was created for $C6$.

The main takeaway here is that $C1$, $C2$, and $C5$ still do not have corresponding server-side values. Our model thus reduced the number of messages sent by the client and the number of arrays created by the server by 3, assuming computing the server side value of each of $C1$, $C2$, and $C5$ involve the client sending one message and the server creating one array. Effectively, delayed evaluation implements dead code elimination, since the values will be computed only when (if) they are needed.

\subsection{Example 3: Common Sub-expression Elimination}
Let's look at the following example to illustrate common sub-expression elimination:
\begin{center}
\begin{tabular}{c}
\begin{lstlisting}
B = A * A
C = A * A 
D = B + C
print(D)
\end{lstlisting}
\end{tabular}
\end{center}

 Assume $A$ points to the $pdarray$ $C1$ which has yet to be computed on the server. $C1$ holds onto a reference to a buffer item. The first line of code would assign a new $pdarray$, $C2$, to $B$ which contains a reference to a buffer item which contains the instructions to multiply $C1$ by itself. A similar process would occur on Line 2, when $C3$ gets assigned to $C$. In the third line of code, a fourth $pdarray$ $C4$ would be created and assigned to $D$. $C4$ would hold a reference to a buffer item that contains the instructions to add $C2$ and $C3$. Finally, the fourth line acts as a trigger to execute the instructions stored by the buffer item that $C4$ references.

The first step in executing that instruction set is resolving any dependencies, which in this case would be on $C2$ and $C3$. First, our model will look at the buffer item that $C2$ references and resolve its dependencies, which would be  $C1$. After sending a message to create $C1$ on the server side and store it as $S1$, our client would then proceed to compute $C2$, which involves multiplying $S1$ by itself. This would result in the creation of $S2$, $C2$'s server value. However, our client will also cache the three address instruction set that the resulting server side array corresponds to. In other words, our client will record that $S2$ is the result of multiplying $S1$ by itself. When computing $C3$'s server side value, the client checks the cache and  realizes that $S1 * S1$ has already been computed, and represents $C3$ as $S2$ on the server side. Finally, after  resolving all dependencies, the client will create a final server side array $S3$, which is the result of adding $S2$ to itself. This corresponds to $C4$. To finish this program, Line 4 will cause the client to query the server for the contents of $S3$ and print that to the user.

By introducing caches of function results, our model reduces the number of messages sent between the server and client and the number of operations computed by the server. 

\section{Experimental Evaluation}
\label{sec:Experiments}

\subsection{Algorithms}

To test the results of optimizations, two algorithms from GraphBlas \cite{GraphBLAS} were ported to Arkouda, on both the Chapel and Python side, so that the results could be compared. All experiments were done on a single shared memory node with a Xeon E3-1220 \cite{IntelXeon} processor.

\subsubsection{Triangle Counting}
The first benchmark is graph triangle counting~\cite{azad2015parallel}, using both sparse and dense matrix representation of the graph. This application  has a high number of basic \textit{pdarray} operations, and  optimizing which is the main goal of this work. Also, it has real world usage~\cite{azad2015parallel}, validating that our optimization will be beneficial to the users of the Arkouda platform.
\paragraph{Dense matrices}
Given a lower triangular matrix $L$ which represents an undirected graph, the number of triangles can be counted by the formula given by \cite{Davis2018}: $$num\_triangles = sum((L*L).*L)$$ Here, $*$ denotes regular matrix multiplication, and $.*$ denotes element-wise matrix multiplication. \\
Since Arkouda does not support two-dimensional arrays on the server side, dense matrices were implemented as an array of \textit{pdarrays} on the client side, so that the Chapel server had no notion of the dimensionality of the problem. Also, regular matrix multiplication, which the Chapel Arkouda server doesn't implement, has been implemented as summation of a vector that is given as a result of element wise multiplication of rows in matrix $L$ and its transpose $L^T$. To be more precise, given matrices $A$ and $B$, their matrix product can be calculated as: $$C(i,j) = sum(A(i)*B^T(j))$$
where $A(i)$ and $B(j)$ denote their respective rows. This maps the problem to the Arkouda APIs.
\paragraph{Sparse matrices}
In this example, we considered sparse matrices in both the CSR (Compressed sparse row) and CSC (Compressed sparse column) format. Since Arkouda already supports set operations, we  use those to count the number of triangles based on the Algorithm~\ref{alg:TriangleCount} which is applied on the neighbourhood matrix of an undirected graph A, requiring both its CSE and CSR form~\cite{GraphBLAS}.

\begin{algorithm}
\begin{footnotesize}
 $(p1, c) \leftarrow CSC(A)$\\
 $(p2, r) \leftarrow CSR(A)$ \\
 $s=0$\\
 \For{i in 1..size(p1)}{
    \For{j in p[i]..p[i+1]}
    {
        $S+=|c[p1[k]..p1[k+1]] \cap r[p2[c[j]]..r[p2[c[j]+1]$ \\
    }
  }
return $S$
\caption{Triangle count for sparse matrices}
\label{alg:TriangleCount}
\end{footnotesize}
\end{algorithm}

Since splices of arrays given above can be represented as sets, we can use the already existing Arkouda intersect operation.
\subsubsection{Betweenness Centrality}
Another important algorithm in graph theory is Betweenness Centrality algorithm, which calculates the \textit{betweenness} measure of a certain node in a graph. This represents the measure of "centrality" of a node, given by the number of shortest paths that go through the node, divided by the number of shortest paths in general in the graph. As per GraphBlas~\cite{GraphBLAS}, the algorithm for returning a betweenness centrality vector from a given node $source$ is given in Algorithm~\ref{alg:BetweennessCentrality}.
\begin{algorithm}
\begin{footnotesize}
 $(n,n)\leftarrow shape(A)$\\
 $delta\leftarrow zeroes(n)$\\
 $sigma\leftarrow zeroes(n,n)$\\
 $q[source]\leftarrow 1$,
 $p\leftarrow q$,
 $d\leftarrow 0, sum\leftarrow 0$\\
 \While{True}{
    $sigma[d]\leftarrow q$\\
    $p \leftarrow p+q$\\
    $q \leftarrow (q*A)*\overline{p}$\\
    $sum = sum(q)$\\
    $d+=1$\\
    \If{s=0} 
    {
        break
    }
  }
\For{i in d-1..0}{
    $t1 \leftarrow 1+delta$\\
    $t2 \leftarrow \frac{t1}{sigma[i]}$\\
    $t3 \leftarrow t2*A$ \\
    $t4 \leftarrow t4*t3$ \\
    $delta \leftarrow delta+t4$
}
return $delta$

\caption{Betweenness Centrality algorithm}
\label{alg:BetweennessCentrality}
\end{footnotesize}
\end{algorithm}
As we can see from Algorithm~\ref{alg:BetweennessCentrality}, the temporary variables $t1, t2, t3, t4$ are overwritten in each iteration of the second loop, and should be amenable to our temporary reuse optimization.
\subsubsection{NYC Taxi Example}
The \textit{NYC Taxi} example is one of the Jupyter notebooks for exploratory data analysis from the Arkouda repository, which consists of Arkouda operations applied to the database of NYC taxi trips in January 2020\cite{ArkoudaGitHub}. This example can be summarized as a series of unary operations applied to a single immutable Arkouda array. The sequence of Arkouda operations is shown in Figure ~\ref{fig:TaxiCab}. As such, it presents an excellent target for memoization of function results. 

\begin{figure}
    \centering
$$min() \rightarrow max() \rightarrow mean() \rightarrow std() \rightarrow min() \rightarrow min()$$
    \caption{Order of operations in the Taxi Cab example}
        \label{fig:TaxiCab}
\vspace{-.1in}
\end{figure}

It is important to point out that some of the redundancies in this example are not apparent to the user. Some Arkouda operations internally have to call other Arkouda operations to complete the task. For example, both $mean()$ and $std()$ (standard deviation) internally call $sum()$. The user would not be able to simply rewrite the Taxi Cab example to avoid all redundant computation. However, our function memoization optimization eliminates these kinds of redundancies as well.
\subsection{Results}
The above mentioned algorithms for Triangle counting and Betweenness Centrality were tested using both dense and sparse matrices in the \textit{SuiteSparse Matrix collection}~\cite{Davis2011}, which provided matrices of varying sizes. The dimensions of each matrix used, as well as the number of non-zeros, is given in Table~\ref{tab:Matrices}.
\begin{table}
    \centering
    \caption{Matrices used in the experiments}
    \begin{tabular}{|l|l|l|}
    \hline
        Size & Non - zeros & Name \\ \hline
        47 x 47 & 472 & mycielskian6 \\ \hline
        62 x 62 & 318 & dolphins  \\ \hline
        105 x 105 & 882 & polbooks \\ \hline
        124 x 124 & 12068 & Journals \\ \hline
        191 x 191 & 4720 & mycielskian8 \\ \hline
        352 x 352 & 458 & GD00\_a  \\ \hline
        366 x 366 & 2440 & dermatology\_5NN  \\ \hline
        400 x 400 & 5656 & Olivetti\_norm\_10 \\ \hline
        453 x 453 & 4065 & celegans\_metabolic  \\ \hline
        492 x 492 & 2834 & Erdos991 \\ \hline
        571 x 571 & 9668 & micromass\_10NN \\ \hline
        1024 x 1024 & 6112 & delaunay\_n10 \\ \hline
        1133 x 1133 & 5451 & Email \\ \hline
        2048 x 2048 & 12254 & delaunay\_n11 \\ \hline
        4096 x 4096 & 24528 & delaunay\_n12 \\ \hline
        8192 x 8192 & 49049 & delaunay\_n13 \\ \hline
        16384 x 16384 & 98244 & delaunay\_n14 \\ \hline
        18772 x 18772 & 396160 & ca-AstroPh \\ \hline
        22499 x 22499 & 87716 & cs4 \\ \hline
        23133 x 23133& 186,936 & cond-mat \\ \hline
        32768 x 32768 & 196548 & delaunay\_n15 \\ \hline
        114599 x 114599 & 239332 & luxembourg\_osm \\ \hline
    \end{tabular}
    \label{tab:Matrices}
\vspace{-.1in}
\end{table}
\subsubsection{Triangle Counting}
\begin{figure}
\centerline{\includegraphics[width=0.6\linewidth]{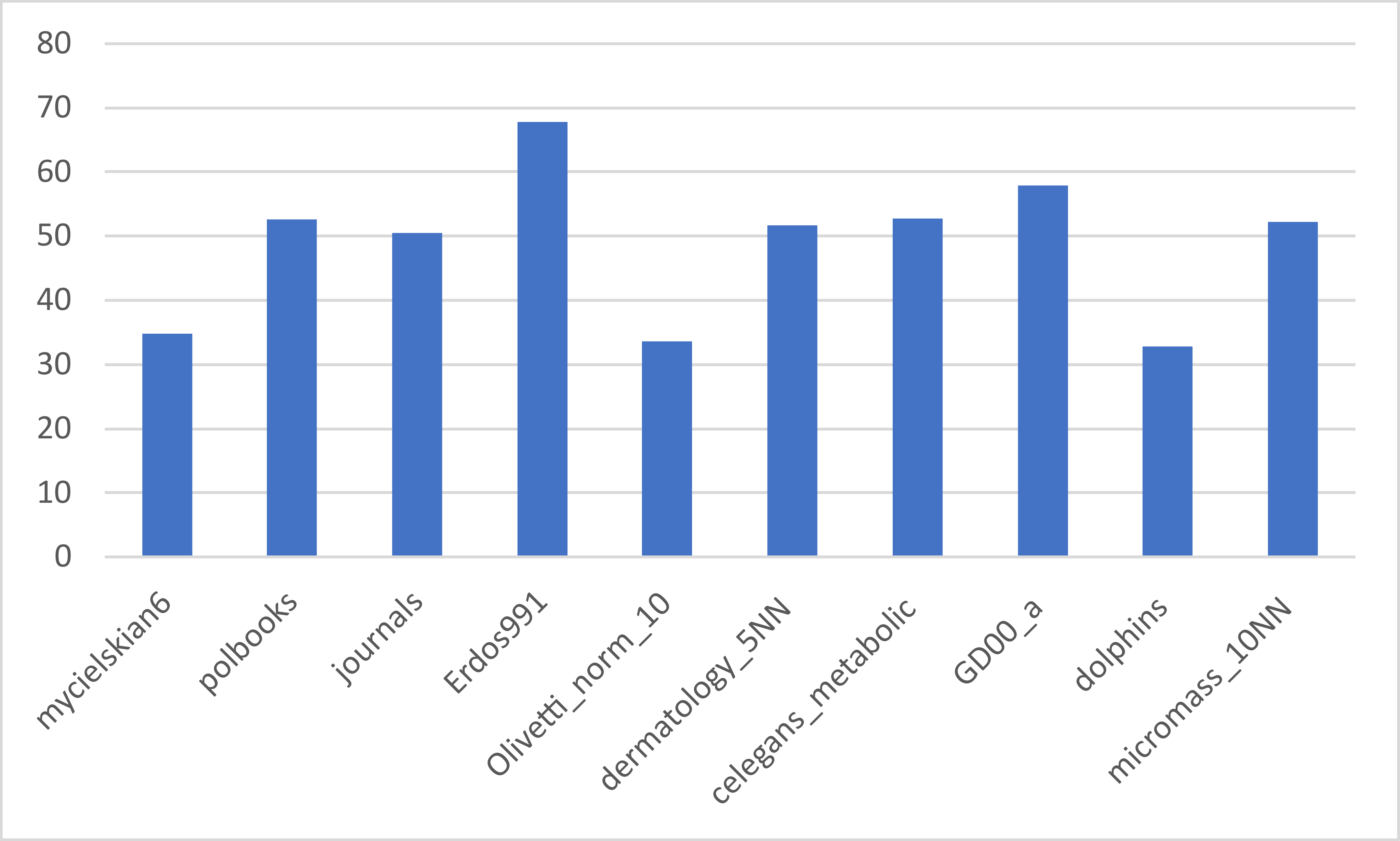}}
\caption{Dense Triangle Count: performance improvements (percentage)}
\label{fig:DenseSpeedup}
\vspace{-.1in}
\end{figure}

\paragraph{Dense matrices}
Matrices of small size (up to $\approx$ 1000 rows) were used in testing triangle count for dense matrices, since the algorithmic complexity of this algorithm is $O(n^3)$. The execution times comparison (speedup) is given in Figure~\ref{fig:DenseSpeedup}.

Since these matrices are dense, the number of non-zeros does not impact the performance. From Figure~\ref{fig:DenseSpeedup}, we can observe consistent improvements between $30\%$ and $60\%$ from our optimizations, across different sizes and number of non-zeros for different matrices. Notably, our improvements are larger for larger matrices.

\begin{figure}
\centerline{\includegraphics[width=0.6\linewidth]{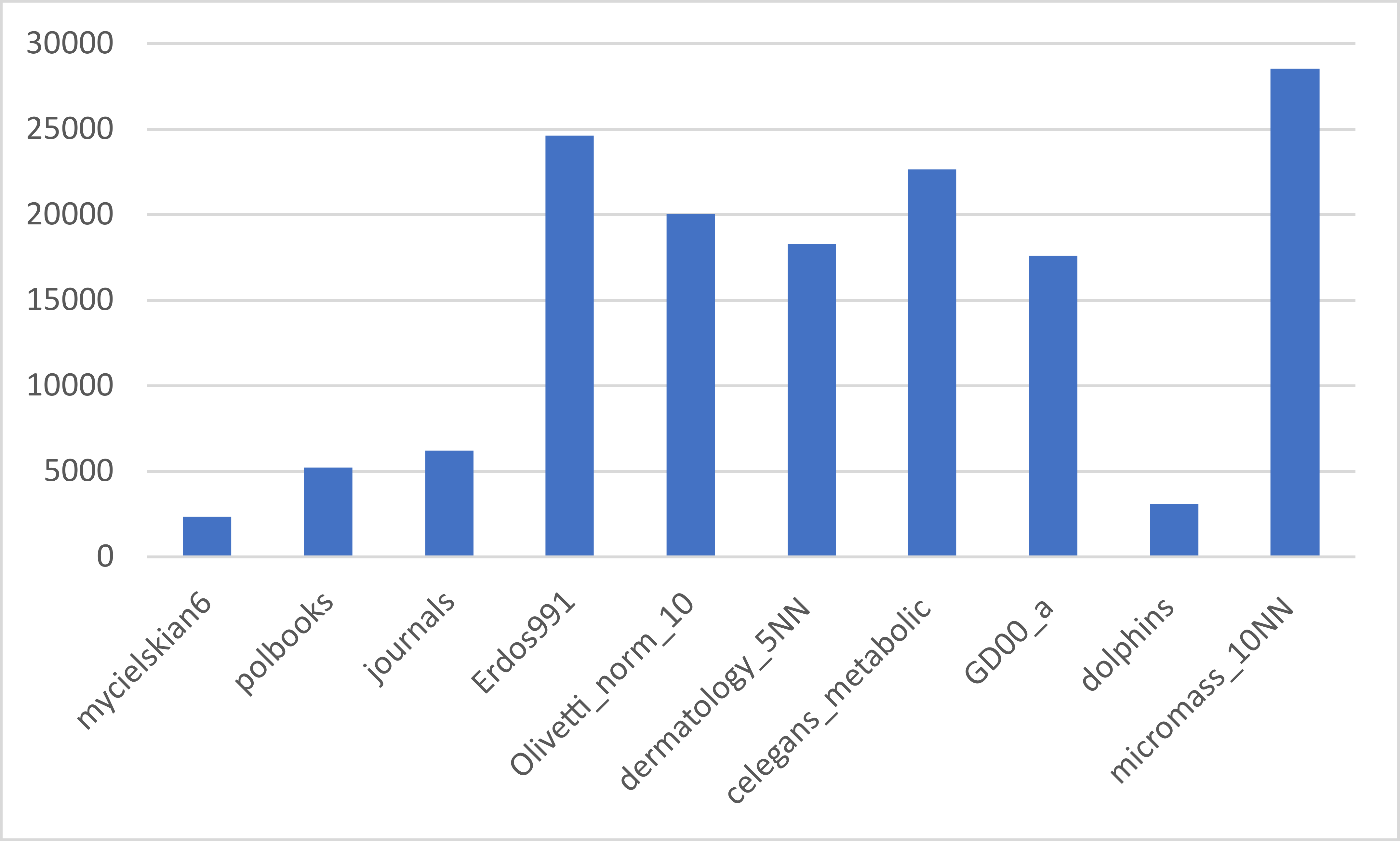}}
\caption{Dense Triangle Count: Ratio of created arrays between base and optimized Arkouda}
\label{fig:DenseRatio}
\vspace{-.1in}
\end{figure}

Since the major part of our optimization focuses on reducing the number of created and destroyed distributed arrays on the Chapel server, the ratio of crated server side arrays between the base and optimized version was also tracked and shown in Figure~\ref{fig:DenseRatio}. We can observe a large reduction in the number of created arrays on the server side for our optimized Arkouda. In base Arkouda, each operation creates a new server side array as a result, and the algorithm has $O(n^3)$ operations. This results in many more arrays created, compared to our optimized version, where temporary arrays are reused. The ratio of created arrays on the server side is larger for lager matrices, since they provide more reuse opportunities.

\begin{figure}
\centerline{\includegraphics[width=0.6\linewidth]{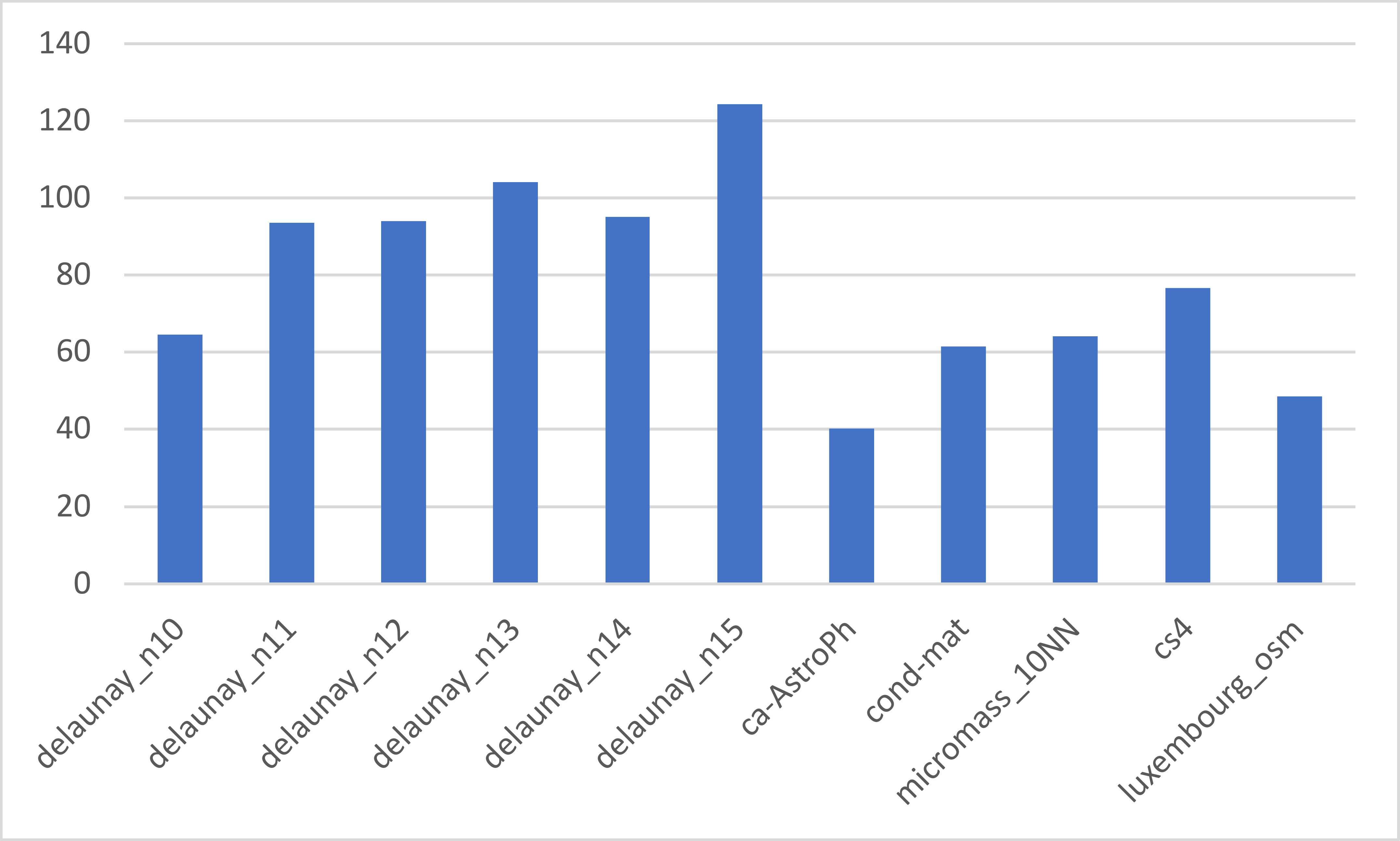}}
\caption{Sparse Triangle Count: performance improvements (percentage)}
\label{fig:SparseSpeedup}
\vspace{-.1in}
\end{figure}

\paragraph{Sparse matrices}
Triangle count for sparse matrices can be applied to larger matrices, and the results of such computations are shown in Figure \ref{fig:SparseSpeedup}.

From these results, we can observe that the speedup is heavily dependant on number of non zeros, since the size of the arrays in CSR format, as well as the sizes of the intersections of sets depend on it. This also correlates with execution time, as the matrices with larger number of non-zeros take more time. Because of the nature of the optimizations, which are reliant on reusing server side arrays of the same size, if there is a large number of non zero-rows, there is a smaller impact of temporary reuse, as the most of execution time is spent on set operation computation. Thus, the more sparse, delaunay matrices, are susceptible to larger performance improvements.

\begin{figure}
\centerline{\includegraphics[width=0.6\linewidth]{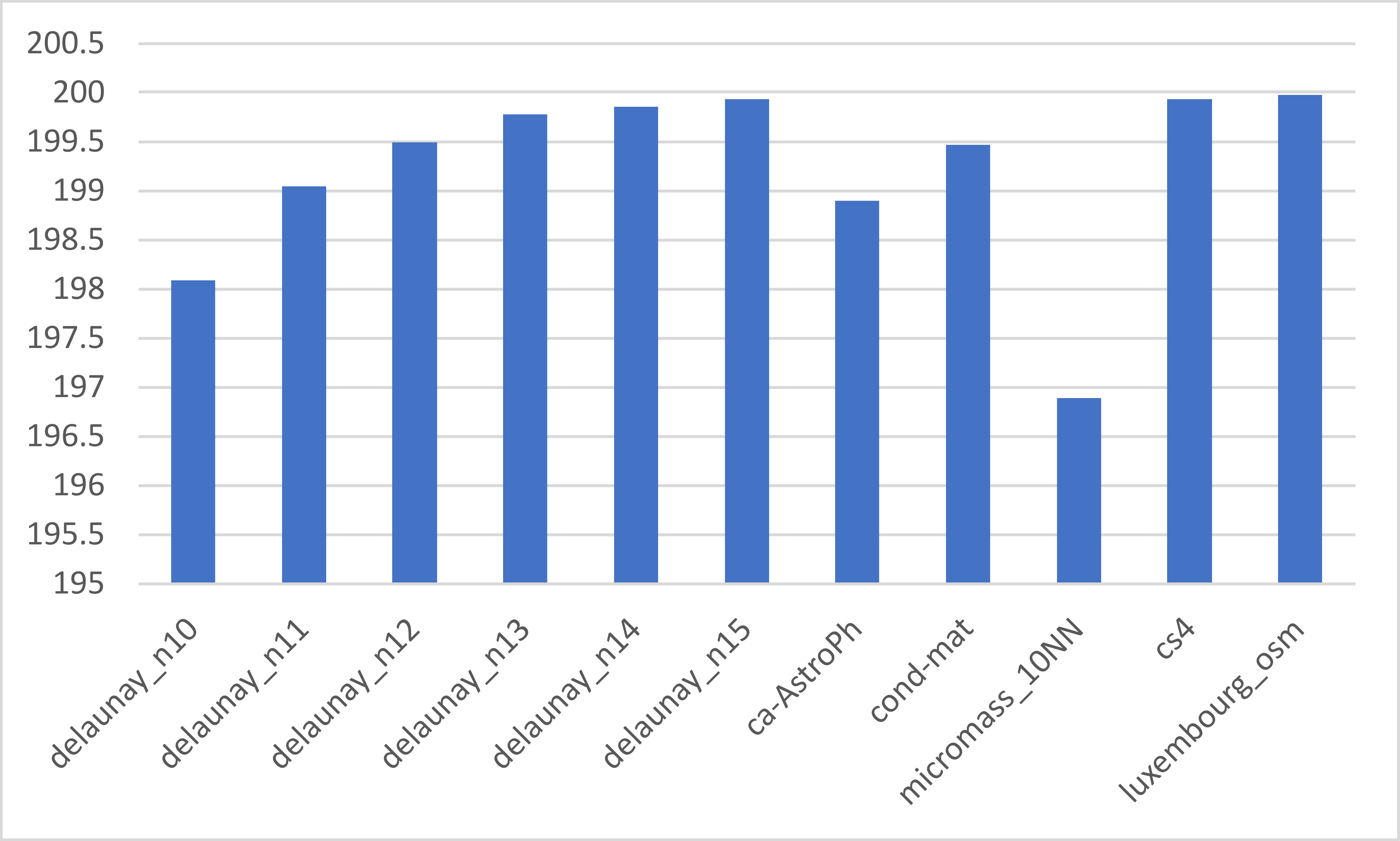}}
\caption{Sparse Triangle Count: Number of created arrays in base Arkouda over optimized Arkouda (percentage)}
\label{fig:SparseRatio}
\vspace{-.1in}
\end{figure}

As before, we tracked the ratio of created arrays on the server side, and the results are given in figure \ref{fig:SparseRatio}.

Here, it is also shown that the number of non-zeros does not have an impact on the number of created arrays. Since most of the reuse of the temporaries happens inside the loop of the Algorithm~\ref{alg:TriangleCount} which represent the  non-zero part of each row in the matrix, there is almost a fixed ratio between the numbers in optimized and base case (approximately 3).  

\subsubsection{Betweenness Centrality}
\begin{figure}
\centerline{\includegraphics[width=0.6\linewidth]{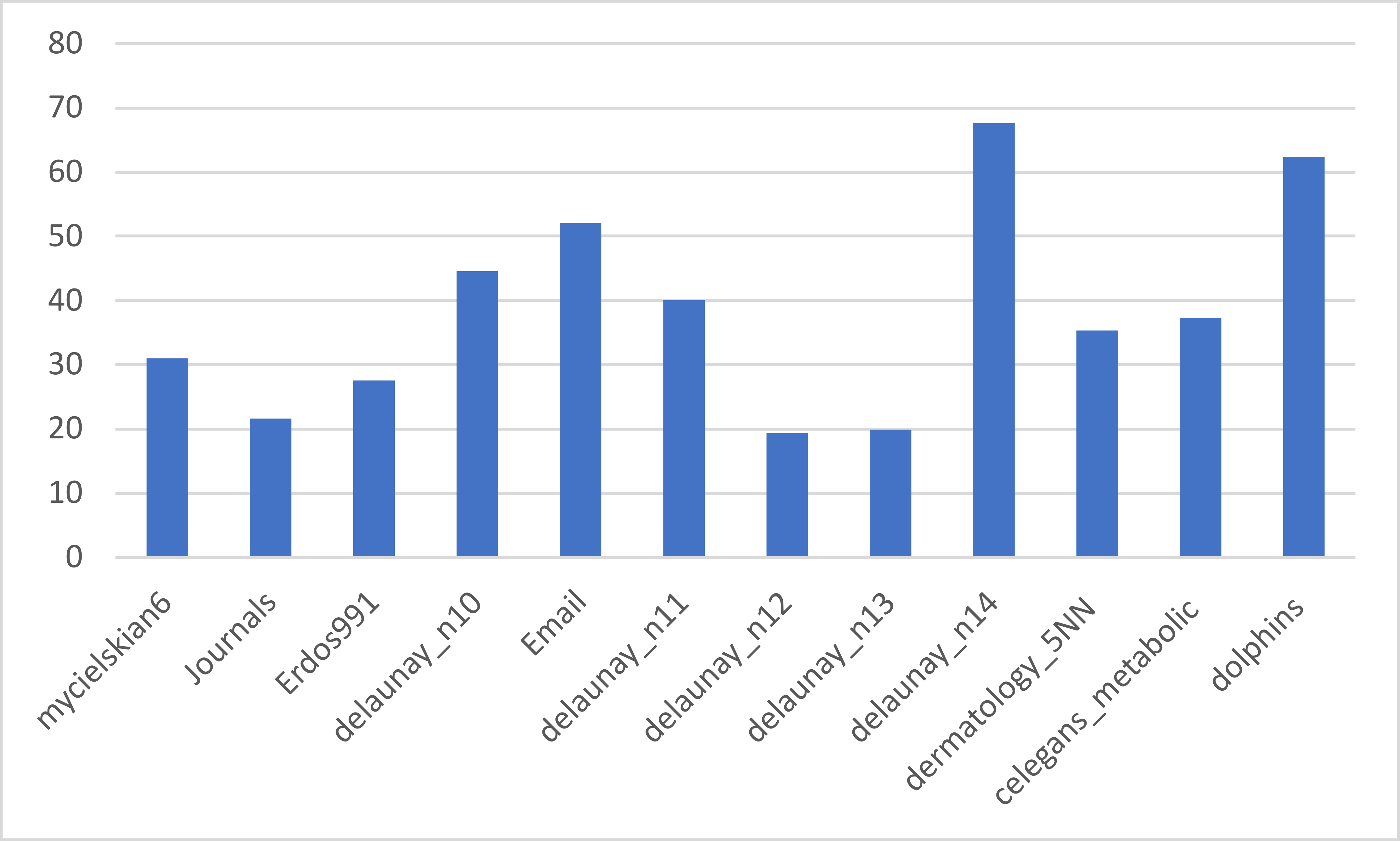}}
\caption{Betweenness Centrality: performance improvements (percentage)}
\label{fig:BetCenSpeedup}
\vspace{-.1in}
\end{figure}

Since the complexity of the Betweenness Centrality algorithm is at most $O(n^2)$, for experiments, larger dense matrices can be used then those for dense triangle count. The times of execution for this algorithm, as well as the speedup, are given in Figure~\ref{fig:BetCenSpeedup}.

From Figure~\ref{fig:BetCenSpeedup} we can derive that the performance improvement is relatively consistent (between $20\%$ $65\%$ and not dependant on the size of the matrix, or the number of non zeros. This is because the opportunities for reusing server side arrays do not change with neither of these parameters, since the size of all arrays that are in the algorithm is the same.

\begin{figure}
\centerline{\includegraphics[width=0.6\linewidth]{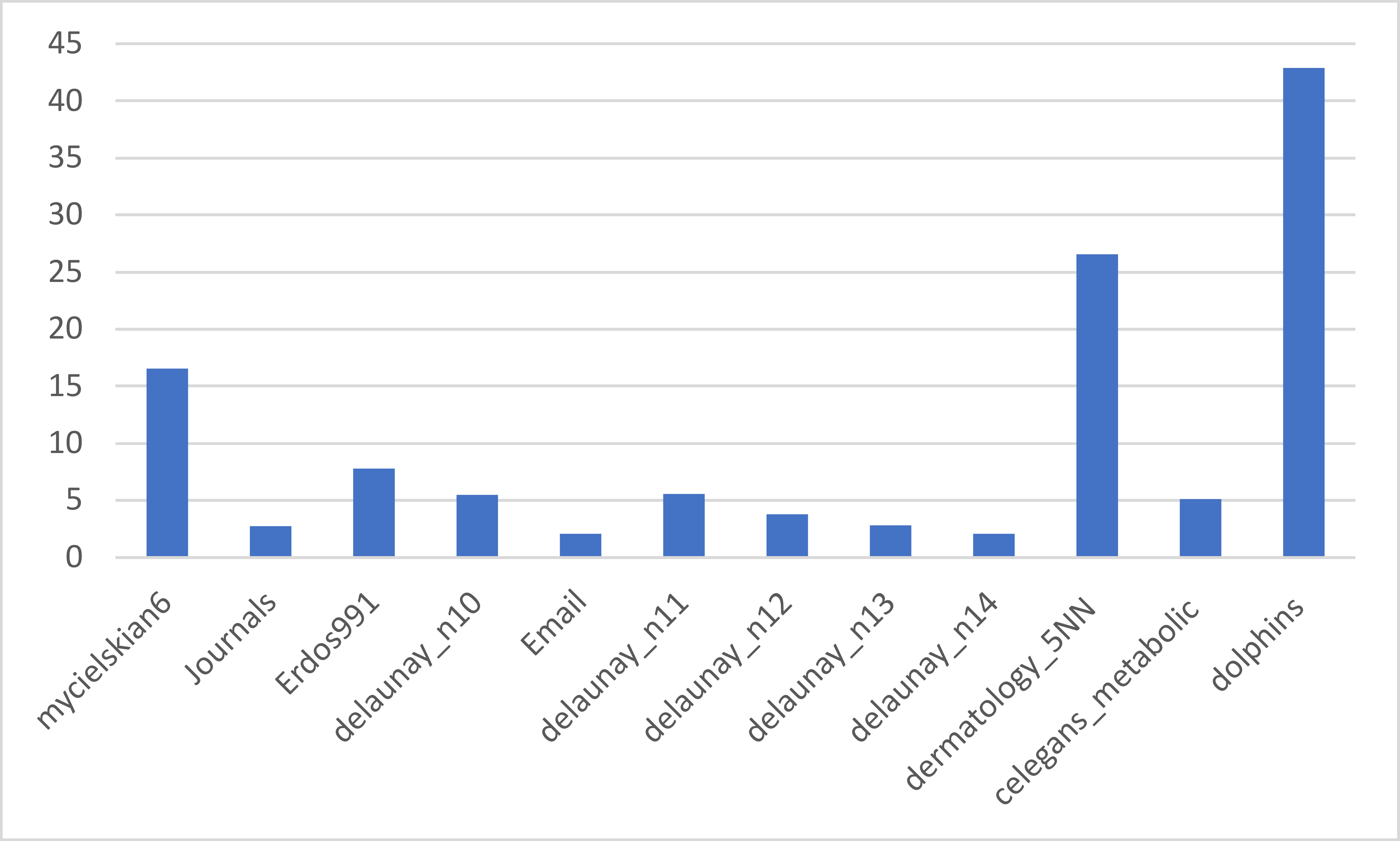}}
\caption{Betweenness Centrality: Number of created arrays in base Arkouda over optimized Arkouda (percentage)}
\label{fig:BetCenRatio}
\vspace{-.1in}
\end{figure}

In Figure~\ref{fig:BetCenRatio} we see that the difference between the number of created server side arrays in base Arkouda and the optimization is not as large as in the triangle count example, simply because there are not as many temporary arrays being created. The bulk of the performance improvements comes from reducing the number of messages that are sent to the server and their serialization and deserialization.
\subsubsection{NYC Taxi Example}
In this example, mostly due to the omission of unnecessary operations, which would calculate already known values, the performance was improved by $35\%$ by both reducing the number of messages to the server, and by avoiding the operations themselves, as shown in Table~\ref{table:TaxiSpeed}.

\begin{table}[H]
    \centering
    \caption{Execution time and the number of sent messages when executing the Taxi Cab example from the Arkouda Notebooks repository}
    \begin{tabular}{|l|l|l|}
    \hline
         & Base [s] & Optimization [s] \\ \hline
        Execution time & 0.25 & 0.16 \\ \hline
        Number of sent messages & 36 & 30 \\ \hline
    \end{tabular}
    \label{table:TaxiSpeed}
\vspace{-.1in}
\end{table}

\subsubsection{Cost breakdown}
\begin{figure}
\centerline{\includegraphics[width=1\linewidth]{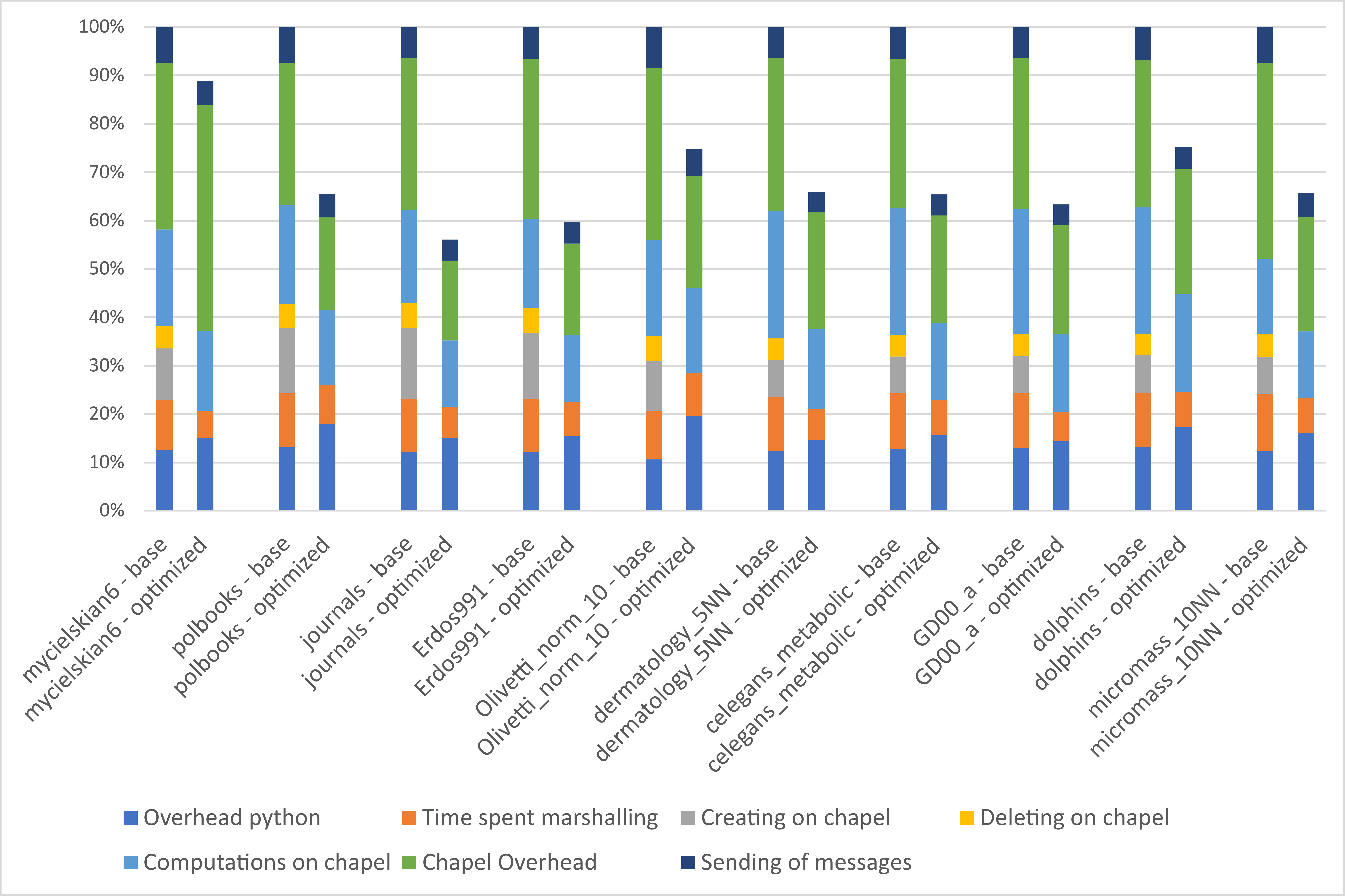}}
\caption{Dense Triangle Count: cost breakdown}
\label{fig:DenseProf}
\vspace{-.1in}
\end{figure}

\begin{figure}
\centerline{\includegraphics[width=1\linewidth]{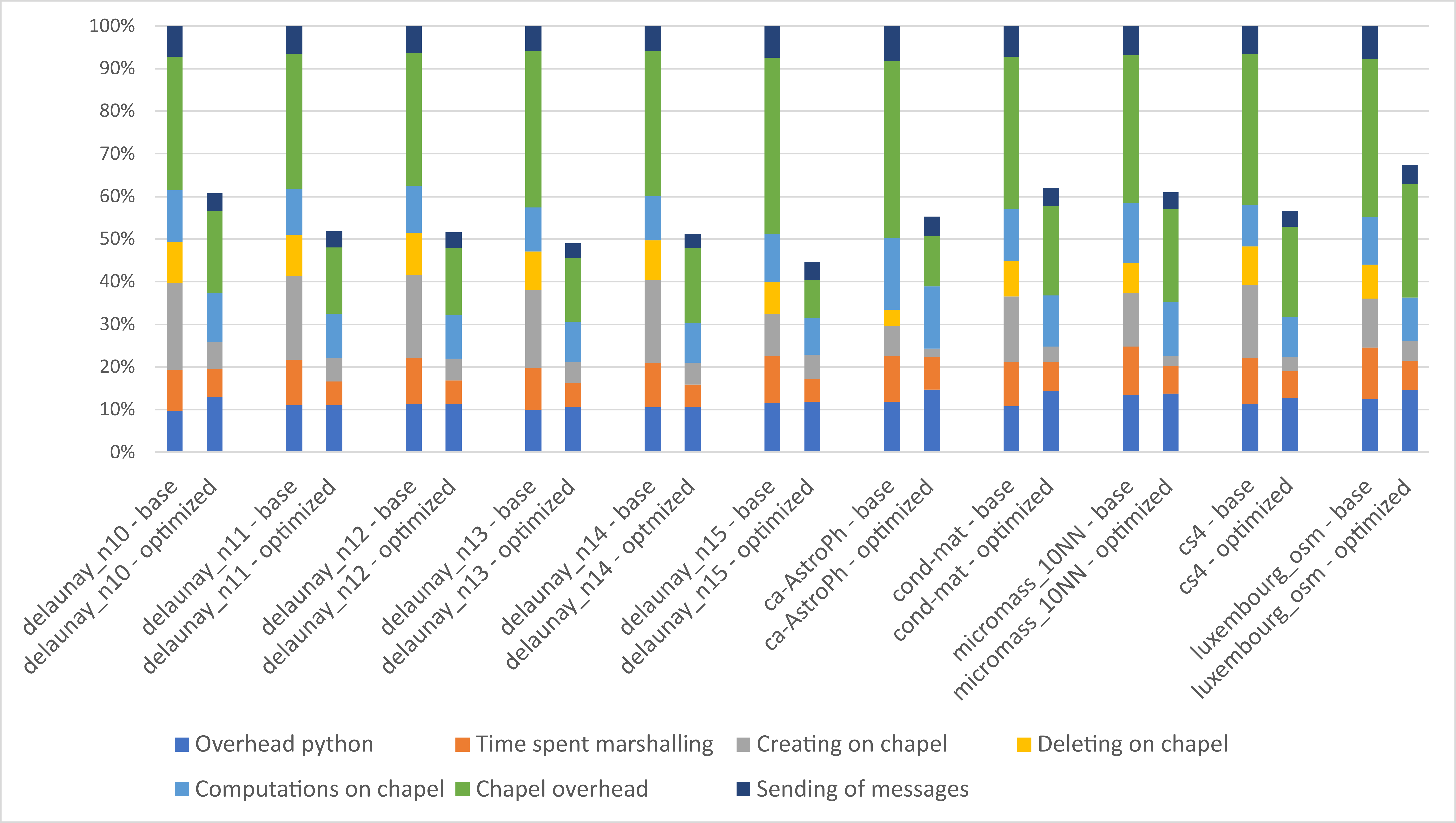}}
\caption{Sparse Triangle Count: cost breakdown}
\label{fig:SparseProf}
\vspace{-.1in}
\end{figure}

\begin{figure}
\centerline{\includegraphics[width=1\linewidth]{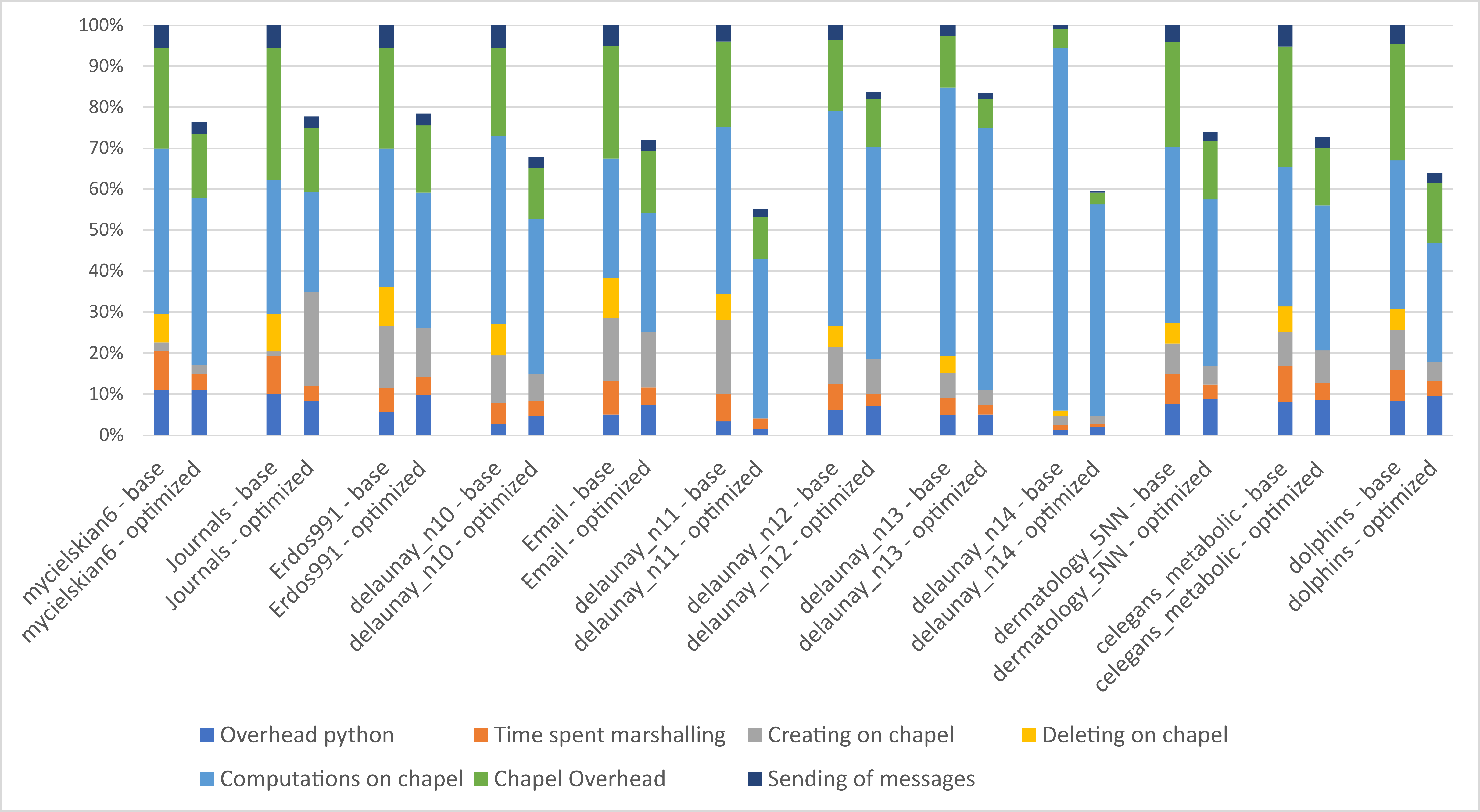}}
\caption{Betweenness Centrality: cost breakdown }
\label{fig:BetCenprof}
\vspace{-.1in}
\end{figure}

To better understand the the nature of our individual optimizations and how they affect the execution time,  we divided the execution into several parts:
\begin{itemize}[topsep=0pt,itemsep=-1ex,partopsep=1ex,parsep=1ex,leftmargin=*]
    \item {\em Overhead Python} - time spent on the setup needed for the Arkouda classes and storing the client side arrays
    \item {\em Time spent marshalling} - time spent on marshalling and unmarshalling the arguments of messages sent and received on the Python side
    \item {\em Creating on Chapel} - time spent creating and adding to the symbol table on the Chapel server
    \item {\em Deleting on Chapel} - time spent deleting the server side arrays on the server
    \item {\em Computations on Chapel} - time spent doing computational work on the Chapel server (adding, multiplying arrays etc.)
    \item {\em Chapel overhead} - time spent on the marshalling and unmarshalling the arguments of messages sent and received on the server side
    \item {\em Sending of messages} - time spent sending messages from the server to the client and vice versa
\end{itemize}
The cost breakdown of the algorithms Dense Triangle Count, Sparse Triangle Count and Betweenness Centrality is shown in figures \ref{fig:DenseProf}, \ref{fig:SparseProf} and \ref{fig:BetCenprof}, respectively.

From Figures \ref{fig:DenseProf}, \ref{fig:SparseProf} and \ref{fig:BetCenprof} we can observe that, as expected, the time saved on creating the arrays is proportional to the ratios given in Figures~\ref{fig:DenseRatio}, ~\ref{fig:SparseRatio}, and ~\ref{fig:BetCenRatio}. As the number of messages is reduced with elimination of delete messages, the time spent marshalling and unmarshalling parameters on both the client and server side decreases significantly. Python overhead increases in optimized Arkouda, as expected, but that increase is negligible compared to the overall decreases of the execution time. We can also observe minor improvement in the computations on the chapel server. We suspect that this comes from the temporary reuse, which in turn lowers the pressure on the data caches on the server.  

\begin{figure}
\vspace{-.1in}
\centerline{\includegraphics[width=0.65\linewidth]{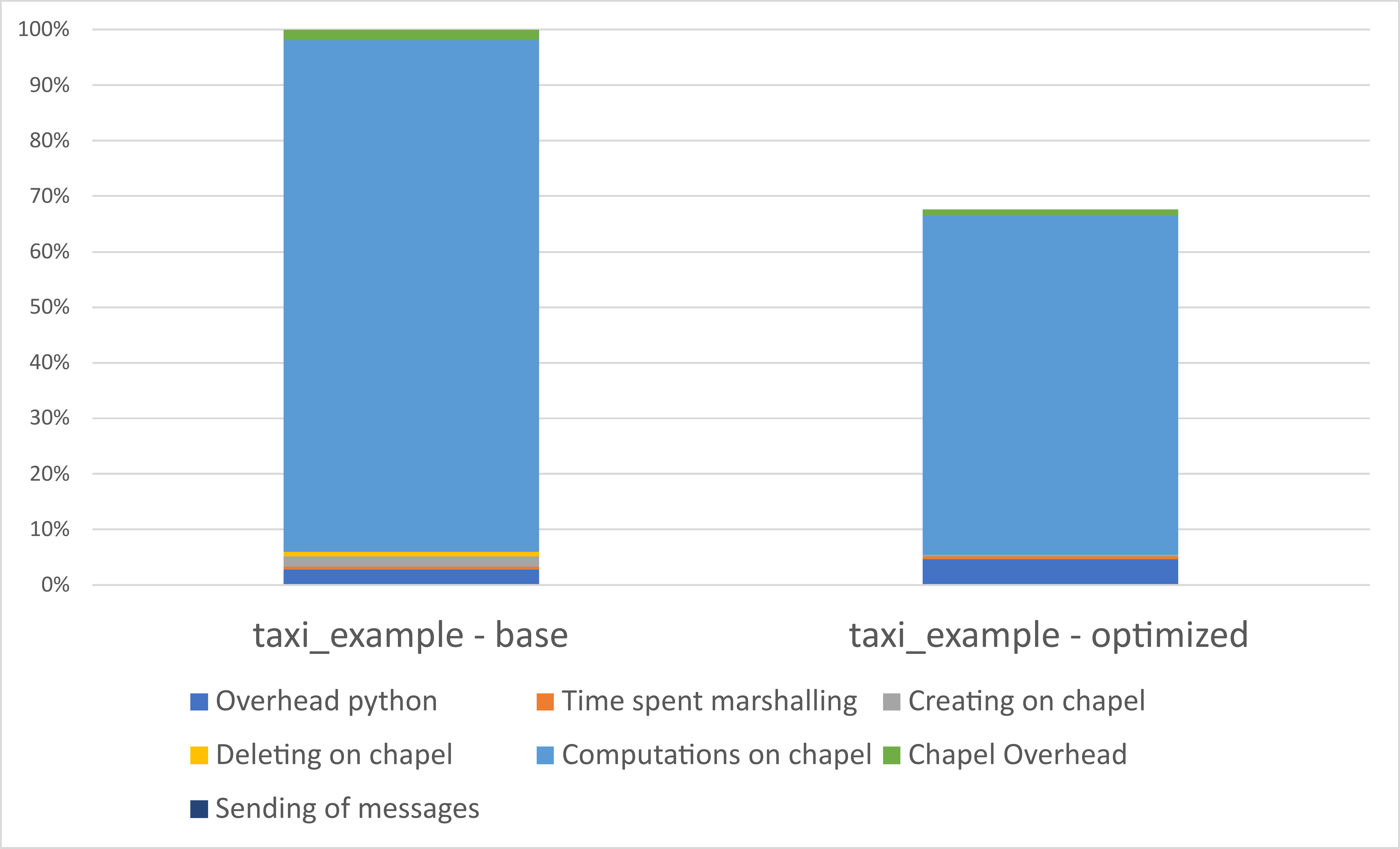}}
\caption{Taxi example : cost breakdown }
\label{fig:TaxiProf}
\vspace{-.1in}
\end{figure}

The impact of memoization on the time of execution can be better understood by  a similar graph on Figure~\ref{fig:TaxiProf}) for the Taxi Cab example.

As there is a small amount of messages that are sent between the client and the server, most of the execution time is spent doing computations on the Chapel server. We can observe that, by doing memoization, we reduced the time spent on that computation significantly, which represents the bulk of the overall time saved.

\section{Future Work}
\label{sec:Future}

For future work, we plan on further reducing the number of messages that the client sends to the server by abstracting away a piece of Python code containing several Arkouda commands into a single Python lambda function. This lambda can be sent as one message to the server which will then be executed on the server side. This approach will serve as a basis to future optimizations, including batching Python loop commands in one message. Determining how to send Python lambdas in messages and interpreting those messages on the Chapel side is a necessary prerequisite to this approach. For example, consider the snippet of code used in the counting sort of large arrays, specifically the one that counts the number of elements that have a certain digit in a certain place (code shown below).

\begin{algorithm}
\begin{footnotesize}
\For{i in 0, len(array)}{
    $b=floor(((array[i]-m)/e)\%r)$\\
    $buckets[b]+=1$
}
\caption{Snippet from the counting sort algorithm}
\label{alg:CountingSort}
\end{footnotesize}
\end{algorithm}

We can abstract the \textit{for} loop in this code, and migrate it to the Chapel server, where it can be parallelized, as opposed to the sequential execution as specified above. The client would send a single message, which would look similar to line of code below, where $array$ and $b$ are the arrays in question, followed by the number of needed parameters for the lambda and the lambda function itself. The last parameter would be a list consisting of all the parameters that the lambda function needs to successfully execute. 
$$count\_buckets(array, b, 3, f, [m,e,r])$$

By using this or a similar API, and generalizing so that it can be used with many different types of \textit{for} loops, it would be easier for he user to write programs that execute in parallel, without the need for deeper knowledge of Chapel loop parallelization mechanisms.

Another feature we plan on adding is the support for asynchronous messages. The current Arkouda framework only supports blocking operations on the client side, which can influence the ability of the client to quickly process lines of code. Through asynchronous message sending, the client can simply send a large series of messages and be notified of results when the server has finished its computation. Additionally, this optimization will maximize the parallel functionality of the server and will allow the server to process iterations of a sequential Python loop containing Arkouda commands in a pipeline fashion. 

\section{Conclusions}
In this paper, we have presented several optimizations for the Arkouda client-server data analysis framework. By intercepting and buffering the Arkouda commands inside of the Python client interpreter, we were able to perform liveness analysis of the Arkouda server-side Chapel distributed arrays, and implement several optimizations, namely temporary reuse, lazy evaluation, common sub-expression elimination, and memoization. Our optimizations significantly reduce the number of temporaries created on the Chapel server, reduce the number of messages sent between the client and the server, and avoid redundant computations on the Chapel server by both caching the results of the reduction operations on the client side, and by tracking the results of the evaluated subexpressions on the server side.

We evaluated our optimizations on several relevant benchmark applications, and on a large number of inputs, and showed significant performance improvements over base Arkouda, between $20\%$ and $120\%$ across the board. All our optimizations still maintain the fully interactive nature of Arkouda as a platform for exploratory data analysis. 
\label{sec:Conclusions}

\section{Acknowledgements}

This work was supported in part by the Big-Data Private-Cloud Research Cyber infrastructure MRI-award funded by NSF under grant CNS-1338099 and by Rice University's Center for Research Computing (CRC). It was also supported by the United States Department of Energy through Lawrence Berkeley National Laboratory.

\bibliographystyle{plain}
\bibliography{references}

\end{document}